\RequirePackage[mathlines]{lineno}
\documentclass[prd,twocolumn,showpacs,amsmath,amssymb]{revtex4-1}
\usepackage{overpic,graphicx}
\usepackage{dcolumn}
\usepackage{bm}
\usepackage{rotating}
\usepackage{subfigure}
\usepackage{color}

\begin{document}

\title{\bf \boldmath
Observation of $D^{0(+)}\to K^0_S\pi^{0(+)}\eta^\prime$
and improved measurement of $D^0\to K^-\pi^+\eta^\prime$}

\author{
M.~Ablikim$^{1}$, M.~N.~Achasov$^{10,d}$, S. ~Ahmed$^{15}$, M.~Albrecht$^{4}$, M.~Alekseev$^{55A,55C}$, A.~Amoroso$^{55A,55C}$, F.~F.~An$^{1}$, Q.~An$^{52,42}$, Y.~Bai$^{41}$, O.~Bakina$^{27}$
% ...
\\
\vspace{0.2cm}
(BESIII Collaboration)\\
\vspace{0.2cm} {\it
$^{1}$ Institute of High Energy Physics, Beijing 100049, People's Republic of China\\
% ...
$^{4}$ Bochum Ruhr-University, D-44780 Bochum, Germany\\
% ...
$^{10}$ G.I. Budker Institute of Nuclear Physics SB RAS (BINP), Novosibirsk 630090, Russia\\
% ...
$^{15}$ Helmholtz Institute Mainz, Johann-Joachim-Becher-Weg 45, D-55099 Mainz, Germany\\
% ...
$^{27}$ Joint Institute for Nuclear Research, 141980 Dubna, Moscow region, Russia\\
% ...
$^{41}$ Southeast University, Nanjing 211100, People's Republic of China\\
$^{42}$ State Key Laboratory of Particle Detection and Electronics, Beijing 100049, Hefei 230026, People's Republic of China\\
% ...
$^{52}$ University of Science and Technology of China, Hefei 230026, People's Republic of China\\
% ...
$^{55}$ (A)University of Turin, I-10125, Turin, Italy; (B)University of Eastern Piedmont, I-15121, Alessandria, Italy; (C)INFN, I-10125, Turin, Italy\\
% ...
$^{d}$ Also at the Novosibirsk State University, Novosibirsk, 630090, Russia\\
% ...
}
\vspace{0.4cm}
}

\begin{abstract}
By analyzing an $e^+e^-$ data sample corresponding to an integrated luminosity of 2.93 fb$^{-1}$ taken at a center-of-mass energy of 3.773~GeV with the BESIII detector, we measure the branching
fractions of the Cabibbo-favored hadronic decays $D^0\to K^-\pi^+\eta^\prime$, $D^0\to K^0_S\pi^0\eta^\prime$, and $D^+\to K^0_S\pi^+\eta^\prime$, which are determined to be $(6.43 \pm 0.15_{\rm stat.} \pm 0.31_{\rm syst.})\times 10^{-3}$, $(2.52 \pm 0.22_{\rm stat.} \pm 0.15_{\rm syst.})\times 10^{-3}$, and $(1.90 \pm 0.17_{\rm stat.} \pm 0.13_{\rm syst.})\times 10^{-3}$, respectively. The precision of the branching fraction of $D^0\to K^-\pi^+\eta^\prime$ is significantly improved, and the processes $D^0\to K^0_S\pi^0\eta^\prime$ and $D^+\to K^0_S\pi^+\eta^\prime$ are observed for the first time.
\end{abstract}

\pacs{13.25.Ft, 14.40.Lb}

\oddsidemargin  -0.2cm
\evensidemargin -0.2cm

\maketitle

\section{Introduction}

Hadronic decays of $D$ mesons provide important information to understand the
weak and strong interactions in the charm sector. Various experiments have measured the branching fractions of
hadronic decays of $D$ mesons~\cite{pdg2014}. However, the measurement accuracy of the Cabibbo-favored (CF) decays $D\to \bar K\pi\eta^\prime$
is still very poor~\cite{pdg2014}. The Particle Data Group (PDG) gives a branching fraction of $(0.75\pm0.19)\%$ for $D^0\to K^-\pi^+\eta^\prime$,
which was measured by the CLEO collaboration 25 years ago~\cite{prd48_4007,pdg2014}. There are no measurements for the isospin-related decay modes
$D^0\to K^0_S\pi^0\eta^\prime$ and $D^+\to K^0_S\pi^+\eta^\prime$. The statistical isospin model (SIM) proposed in Refs.~\cite{rosner1,rosner2}
predicts a simple ratio of the branching fractions for the isospin multiplets: ${\mathcal B}(D^0\to K^-\pi^+\eta^\prime):{\mathcal B}(D^0\to K^0_S\pi^0\eta^\prime):{\mathcal B}(D^+\to K^0_S\pi^+\eta^\prime)
\equiv 1:{\mathcal R^0}:{\mathcal R^+}
\equiv 1:\frac{{\mathcal B}(D^0\to K^0_S\pi^0\eta^\prime)}{{\mathcal B}(D^0\to K^-\pi^+\eta^\prime)}:\frac{{\mathcal B}(D^+\to K^0_S\pi^+\eta^\prime)}{{\mathcal B}(D^0\to K^-\pi^+\eta^\prime)}=1:0.4:0.9$. Precision measurements of the branching fractions of $D\to \bar K\pi\eta^\prime$ are crucial to test the SIM prediction.

In this paper, we report an improved measurement of the branching fraction
for $D^0\to K^-\pi^+\eta^\prime$ and
the first measurements of the branching fractions for $D^0\to K^0_S\pi^0\eta^\prime$ and
$D^+\to K^0_S\pi^+\eta^\prime$.
The analysis is performed using an $e^+e^-$ annihilation data sample
corresponding to an integrated luminosity of $2.93$~fb$^{-1}$~\cite{lum}
collected with the BESIII detector~\cite{bes3} at $\sqrt s=3.773$~GeV.
At this energy,
relatively clean $D^0$ and $D^+$ meson samples are obtained
from the processes $e^+e^-\to\psi(3770)\to D^0\bar D^0$ or $D^+D^-$. To improve statistics,
we use a single-tag method, in which either a $D$ or $\bar D$ is reconstructed in an event.
Throughout the text, charge conjugated modes are implied, and
$D\bar D$ refers to $D^0\bar D^0$ and $D^+D^-$ unless stated explicitly.

\section{BESIII detector and Monte Carlo simulation}
\label{sec:detector}

The BESIII detector is a magnetic spectrometer
that operates at the BEPCII collider.  It has a cylindrical geometry with a solid-angle
coverage of 93\% of $4\pi$. It consists of several main components. A 43-layer
main drift chamber~(MDC) surrounding the beam pipe performs precise
determinations of charged particle trajectories and measures the
specific ionization energy loss~(${\rm d}E/{\rm d}x$) for charged particle identification~(PID).
An array of time-of-flight counters~(TOF) is located outside the MDC and provides additional PID
information. A CsI(Tl) electromagnetic calorimeter~(EMC) surrounds
the TOF and is used to measure the deposited energies of photons and
electrons. A solenoidal superconducting magnet outside the EMC
provides a 1 T magnetic field in the central tracking region of the
detector. The iron flux return of the magnet is instrumented with the resistive plate muon counters arranged in
nine layers in the barrel and eight layers in the endcaps for identification of
muons with momenta greater than 0.5\,GeV/$c$. More
details about the BESIII detector are described in Ref.~\cite{bes3}.

A Monte Carlo~(MC) simulation software package, based on {\sc geant4}~\cite{geant4},
includes the geometric description and response of the
detector and is used to determine the detection efficiency and
to estimate backgrounds for each decay mode.
An inclusive MC sample, which includes the $D^0\bar D^0$, $D^+D^-$ and non-$D\bar D$ decays of the
$\psi(3770)$, initial-state-radiation~(ISR) production of the
$\psi(3686)$ and $J/\psi$, the continuum process $e^+e^-\to q\bar q$~($q=u$,~$d$,~$s$), Bhabha scattering events, di-muon events and
di-tau events, is produced at $\sqrt s=3.773\,{\rm GeV}$.
The equivalent luminosity of the inclusive MC sample is ten times that of the data sample.
The $\psi(3770)$ decays are generated with the MC generator {\sc kkmc}~\cite{kkmc}, which
incorporates the effects of ISR~\cite{isr}.  Final-state-radiation\,(FSR)
effects are simulated with the \textsc{photos} package~\cite{photons}. The known decay modes are generated using
{\sc evtgen}~\cite{evtgen} with branching fractions taken from
the PDG~\cite{pdg2014}, while the remaining unknown decays are
generated using {\sc lundcharm}~\cite{lundcharm}.

\section{Event selection}
\label{sec:evtsel}
In this analysis, all charged tracks are required to be within
$|\rm{cos\theta}|<0.93$,
where $\theta$ is the polar angle with respect to the
positron beam.
Good charged tracks, except those used to reconstruct $K^0_{S}$
mesons, are required to originate from the interaction region defined by
$V_{xy}< 1$~cm and $|V_{z}|< 10$~cm,
where $V_{xy}$ and $|V_{z}|$ are the distances of the closest approach
of the reconstructed tracks to the interaction point (IP), perpendicular to
and along the beam direction, respectively.

Charged kaons and pions are
identified using the ${\rm d}E/{\rm d}x$ and TOF measurements.
The combined confidence levels
for the kaon and pion hypotheses ($CL_{K}$ and $CL_{\pi}$) are calculated and
the charged track is identified as kaon (pion) if $CL_{K(\pi)}$ is greater than
$CL_{\pi(K)}$.

\begin{figure}[htbp]
  \centering
  \includegraphics[width=3.3in]{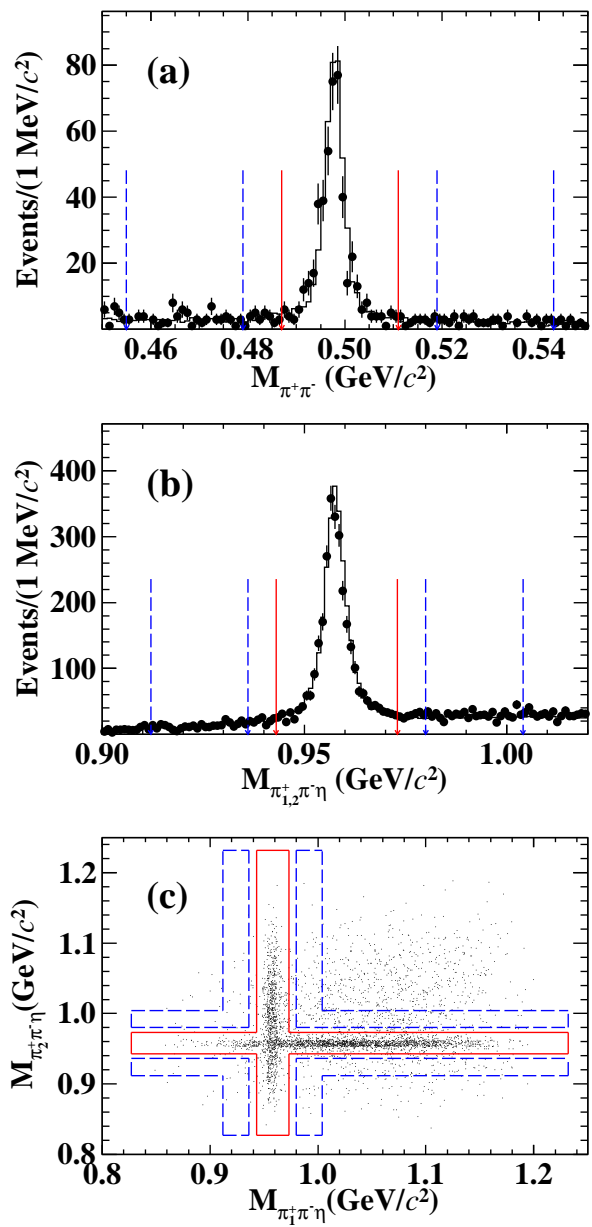}
  \caption{ (Color online)\
(a)~Distribution of $M_{\pi^+\pi^-}$ for the $K^0_S$ candidates from $D^0\to K^0_S\pi^0\eta^\prime$ decays
and (b)~the combined $M_{\pi_{1}^+\pi^-\eta}$ and $M_{\pi_{2}^+\pi^-\eta}$ distribution for the $\eta^\prime$ candidates from $D^0\to K^-\pi^+\eta^\prime$
decays,
where the dots with error bars are data, the histograms are inclusive MC samples,
and the pairs of red solid~(blue dashed) arrows show the
boundaries of the $K^0_S$ or $\eta^\prime$ 1D signal~(sideband) region.
(c)~Scatter plot of $M_{\pi^+_2\pi^-\eta}$ versus $M_{\pi^+_1\pi^-\eta}$ for the $D^0\to K^-\pi^+\eta^\prime$  candidate events in the data sample, where
the range surrounded by the red solid~(blue dashed) lines denotes the $\eta^\prime$ 2D signal~(sideband) region.
In these figures,
except for the $K^0_S$ or $\eta^\prime$ mass requirement,
all selection criteria and an additional requirement of $|M_{\rm BC}-M_D|<0.005$~GeV/$c^2$ have been imposed.
The signal and sideband regions, illustrated here, are applied for all decays of interest in the analysis.
}\label{fig:sig}
\end{figure}

The neutral kaon is reconstructed via the $K^0_S\to\pi^{+}\pi^{-}$ decay mode.
Two oppositely charged tracks with $|V_{z}|< 20$~cm are assumed to be a $\pi^+\pi^-$
pair without PID requirements
and the $\pi^+\pi^-$ pair is constrained to originate from a common vertex.
The $\pi^+\pi^-$ combination with an invariant mass $M_{\pi^+\pi^-}$ in the range
$|M_{\pi^+\pi^-}-M_{K^0_S}|<0.012$\,GeV/$c^2$, where $M_{K^0_S}$ is the nominal $K^0_{S}$ mass~\cite{pdg2014},
and
a measured flight distance from the IP greater than twice its resolution
is accepted as a $K^0_S$ candidate.
Figure~\ref{fig:sig}(a) shows the $\pi^+\pi^-$ invariant mass distribution, where the two solid arrows denote the $K^0_S$
signal region.

Photon candidates are selected using the EMC information.
The time of the candidate shower must be within 700\,ns of the event start time
and the shower energy should be greater than 25\,(50)\,MeV
if the crystal with the maximum deposited energy for the cluster of interest
is in the barrel~(endcap) region~\cite{bes3}.
The opening angle between the candidate shower and
any charged track is required to be greater than $10^{\circ}$
to eliminate showers associated with charged tracks.
Both $\pi^0$ and $\eta$ mesons are reconstructed via the $\gamma\gamma$ decay mode.
The $\gamma\gamma$ combination with an invariant mass within $(0.115,\,0.150)$
or $(0.515,\,0.570)$\,GeV$/c^{2}$ is regarded as a $\pi^0$ or $\eta$ candidate, respectively.
To improve resolution, a one constraint (1-C) kinematic fit is applied
to constrain the invariant mass of the photon pair to
the nominal $\pi^{0}$ or $\eta$ invariant mass~\cite{pdg2014}.

The $\eta^\prime$ mesons are reconstructed through the decay $\eta^\prime\to\pi^{+}\pi^{-}\eta$.
The invariant mass of the $\pi^{+}\pi^{-}\eta$ combination $M_{\pi^{+}\pi^{-}\eta}$
is required to satisfy $|M_{\pi^+\pi^-\eta}-M_{\eta^\prime}|<0.015$\,GeV/$c^2$,
where $M_{\eta^\prime}$ is the nominal $\eta^\prime$ mass~\cite{pdg2014}.
The boundaries of the one dimensional (1D) $\eta'$ signal region are illustrated by the two solid arrows shown in Fig.~\ref{fig:sig}(b).
The $D^{0(+)}\to K^-(K^0_S)\pi^+\eta^\prime$ decay is selected from the $K^-(K^0_S)\pi^+\pi^+\pi^-\eta$ combination.
Since the two $\pi^+$s in the event have low momenta and are indistinguishable,
the $\eta^\prime$ may be formed from either of the $\pi^+\pi^-\eta$ combinations, whose invariant masses
are denoted as $M_{\pi_1^{+}\pi^{-}\eta}$ and $M_{\pi_2^{+}\pi^{-}\eta}$.
Figure~\ref{fig:sig}(c) shows the scatter plot of
$M_{\pi_2^{+}\pi^{-}\eta}$ versus $M_{\pi_1^{+}\pi^{-}\eta}$
for the $D^0\to K^-\pi^+\eta^\prime$ candidate events in the data sample.
Events with at least one $\pi^+\pi^-\eta$ combination in the two dimensional (2D) $\eta^\prime$ signal region,
shown by the solid lines in Fig.~\ref{fig:sig}(c), are kept for further analysis.

To distinguish $D$ mesons from backgrounds, we define two kinematic variables,
the energy difference $\Delta E \equiv E_D-E_{\rm beam}$ and
the beam-constrained mass
$M_{\rm BC} \equiv \sqrt{E^{2}_{\rm beam}-|\vec{p}_{D}|^{2}}$,
where $E_D$ and $\vec{p}_{D}$ are the energy and momentum
of the $D$ candidate in the $e^+e^-$ center-of-mass system and $E_{\rm beam}$ is the beam energy.
For each signal decay mode, only the combination with
the minimum $|\Delta E|$ is kept if more than one candidate passes
the selection requirements.
Mode-dependent $\Delta E$ requirements, as listed in Table~\ref{tab:singletagN_MC},
are applied to suppress combinatorial backgrounds.
These requirements are about $\pm 3.5\sigma_{\Delta E}$ around the fitted $\Delta E$ peaks, where $\sigma_{\Delta E}$ is the resolution of the $\Delta E$
distribution obtained from fits to the data sample.

\section{Data analysis} \label{sec:ana}
\begin{figure}[htbp]
  \centering
  \includegraphics[width=3.3in]{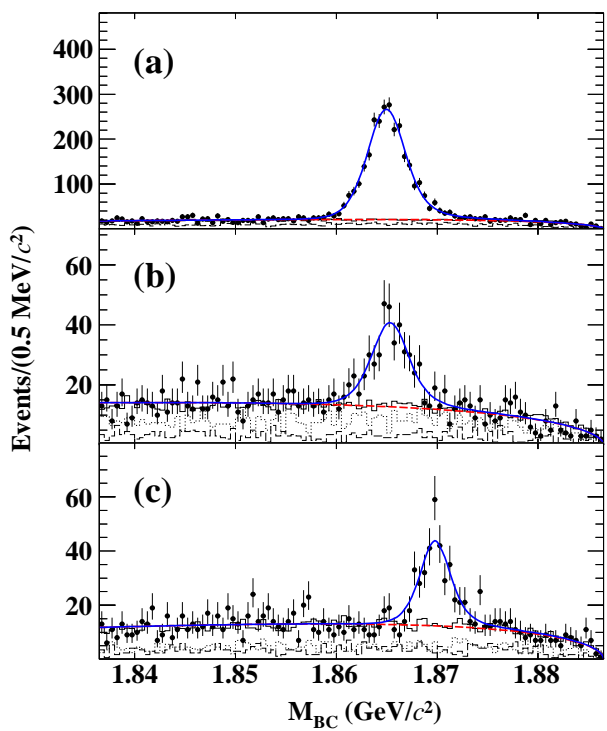}
\caption{(Color online) Fits to the $M_{\rm BC}$ distributions of the (a)
$D^0\to K^-\pi^+\eta^\prime$, (b) $D^0\to K^0_S\pi^0\eta^\prime$, and
(c) $D^+\to K^0_S\pi^+\eta^\prime$ candidate events.
The dots with error bars are data,
the blue solid curves are the total fits
and the red dashed curves are the fitted backgrounds.
The dotted, dashed and solid histograms are
the scaled BKGI, BKGII, and BKGIII components~(see the last paragraph of Sec.~\ref{sec:evtsel}),
respectively.}\label{fig:datafit_Massbc}
\end{figure}

\begin{figure*}[htbp]
\centering
\includegraphics[width=7in]{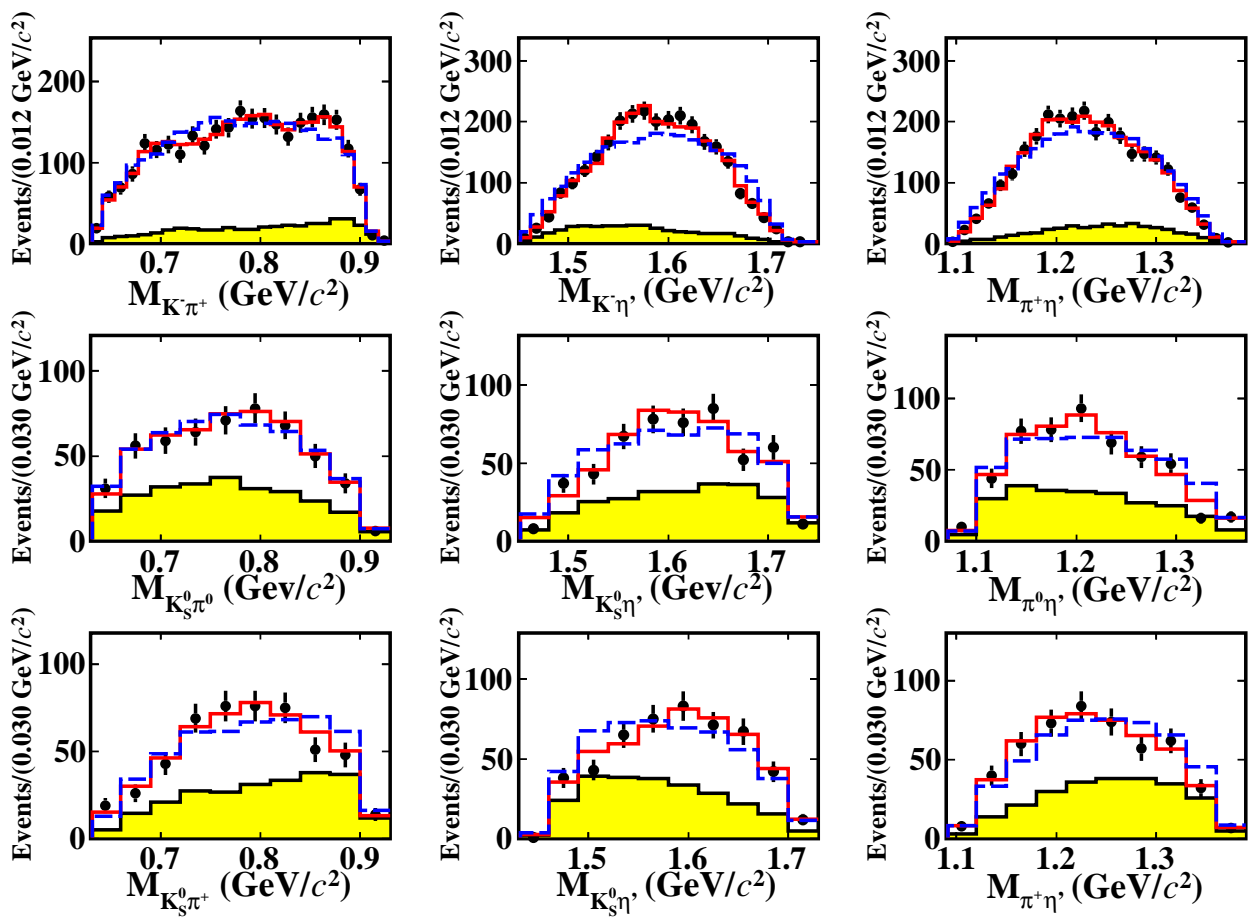}
\caption{
(Color online)
The $M_{K\pi}$,
$M_{\pi\eta^\prime}$, and $M_{K\eta^\prime}$ distributions of data (dots with error bars)
and MC simulations (histograms).
The top, middle, and bottom rows correspond to
$D^0\to K^-\pi^+\eta^\prime$, $D^0\to K^0_S\pi^0\eta^\prime$, and
$D^+\to K^0_S\pi^+\eta^\prime$ candidate events, respectively.
The blue dashed histograms are PHSP MC samples.
The red solid histograms are the modified MC samples.
The yellow shaded histograms are the backgrounds estimated from the inclusive MC sample.
An additional requirement of
$|M_{\rm BC}-M_D|<0.005$~GeV/$c^2$ has been imposed on the events shown in these plots.}\label{fig:mm_compare}
\end{figure*}

The $M_{\rm BC}$ distributions of the accepted candidate events for
the decays of interest in the data sample are shown in Fig.~\ref{fig:datafit_Massbc}.
Unbinned maximum likelihood fits to these spectra are performed
to obtain the $D$ signal yields.
In the fits, the $D$ signal is modeled by an MC-simulated shape convolved with a Gaussian function
with free parameters accounting for the difference between the detector resolution
of the data and that of the MC simulation.
The background shape is described by an ARGUS function~\cite{ARGUS}.
The potential peaking backgrounds are investigated as follows.
The combinatorial $\pi^+\pi^-$~(called BKGI) or $\pi^+\pi^-\eta$~(called BKGII)
pairs in the $K^0_S$ or $\eta^\prime$ signal region may survive
the event selection criteria and form peaking backgrounds around
the $D$ mass in the $M_{\rm BC}$ distributions.
These background components are validated by
the data events in the $K^0_S$($\eta^\prime$) sideband region
defined as $0.020\,(0.022)<|M_{\pi^+\pi^-\,(\pi^+\pi^-\eta)}-M_{K^0_S\,(\eta^\prime)}|<0.044\,(0.046)$\,GeV/$c^2$,
as indicated by the ranges between the adjacent pair of blue dashed arrows
in Fig.~\ref{fig:sig}(a)[(b)].
For $D^0\to K^-\pi^+\eta^\prime$ and $D^+\to K^0_S\pi^+\eta^\prime$ decays,
the data events in the $\eta^\prime$ 2D sideband region,
enclosed by the blue dashed lines in Fig.~\ref{fig:sig}(c), are examined.
For these events, either $M_{\pi_1^{+}\pi^{-}\eta}$ or $M_{\pi_2^{+}\pi^{-}\eta}$
is in the $\eta^\prime$ 1D sideband region, but
both are outside the $\eta^\prime$ 1D signal region.
These two background components are normalized by the ratios of the magnitude of the backgrounds
in the $K^0_S$\,($\eta^\prime$) signal and sideband regions.
The background components from other processes~(called BKGIII) are estimated by analyzing the inclusive MC sample.
The scaled $M_{\rm BC}$ distributions of the surviving events for the
BKGI, BKGII and BKGIII components are shown as
the dotted, dashed and solid histograms in Fig.~\ref{fig:datafit_Massbc}, respectively.
In these spectra, no peaking backgrounds are found,
which indicates that the background shape is well modeled by the ARGUS function.
From each fit, we obtain the number of $D\to \bar K\pi\eta^\prime$ signal events $N_{\rm tag}$,
as summarized in Table~\ref{tab:singletagN_MC}.
The statistical significances of these decays,
which are estimated from
the likelihood difference between the fits with and without
the signal component, are all greater than $10\sigma$.

\begin{figure}[htbp]
\centering
\includegraphics[width=3.4in]{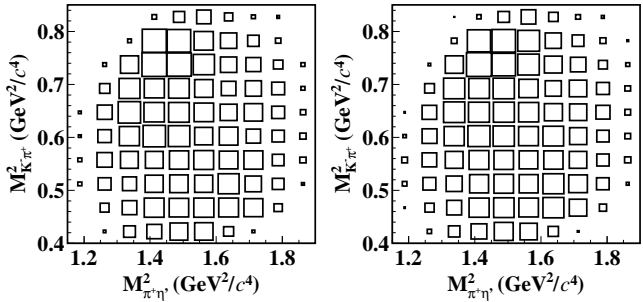}
\caption{\small
Dalitz plots of $M^2_{K^-\pi^+}$ vs. $M^2_{\pi^+\eta^\prime}$
for $D^0\to K^-\pi^+\eta^\prime$ candidate events in data (left) and modified MC sample (right). }\label{fig:dalitz}
\end{figure}

Figure~\ref{fig:mm_compare} shows the $M_{K\pi}$,
$M_{\pi\eta^\prime}$, and $M_{K\eta^\prime}$ distributions
of $D\to \bar K\pi\eta^\prime$ candidate events for data and MC simulations after requiring $|M_{\rm BC}-M_D|<0.005$~GeV/$c^2$.
No obvious sub-resonances have been observed in these invariant mass distributions.
Nevertheless, the phase space (PHSP) MC distributions are not in good agreement with the data distribution (see the blue dashed histograms and dots with errors in Fig.~\ref{fig:mm_compare}).
To solve this problem, we modify the MC generator to produce the correct invariant mass distributions according to the Dalitz plot distributions in data.
In the Dalitz plot, the background component is modeled by the inclusive MC simulation, while the signal component is generated according to efficiency-corrected PHSP MC simulation. In Fig.~\ref{fig:dalitz}, we show the Dalitz plots of $D^0\to K^-\pi^+\eta^\prime$ candidate events for data and the modified MC sample. The invariant mass distributions $M_{K\pi}$, $M_{\pi\eta^\prime}$, and $M_{K\eta^\prime}$ of the modified MC samples are in good agreement with the data distributions (see the red solid histograms and dots with errors in Fig.~\ref{fig:mm_compare}). In the following, we use the modified MC sample to determine the detection efficiencies in the calculation of the branching fractions.

\section{Branching fractions}

The branching fraction of $D\to \bar K\pi\eta^\prime$ is determined according to
 \begin{equation}
 \label{equ:branchingfraction}
 {\mathcal B}(D\to \bar K\pi\eta^\prime) = \frac{N_{\rm tag{\color{blue}}}}{2\cdot N_{D\bar D}\cdot\epsilon
 \cdot{\mathcal B_{\eta^\prime}}
 \cdot{\mathcal B_{\eta}}
 (\cdot{\mathcal B_{\rm inter}})
},
 \end{equation}
where $N_{\rm tag}$ is the number of $D\to \bar K\pi\eta^\prime$ signal events,
$N_{D\bar D}$ is the total number of $D\bar D$ pairs,
$\epsilon$ is the detection efficiency which has been corrected by the differences in the
efficiencies for charged particle tracking and PID, as well as $\pi^0$ and $\eta$ reconstruction, between the data and MC simulation as discussed in Sec.~\ref{sec:ana}, and summarized in Table~\ref{tab:singletagN_MC}.
In Eq.~(\ref{equ:branchingfraction}), ${\mathcal B_{\rm inter}}$ is the product branching fraction $\mathcal B_{K^0_S}\cdot{\mathcal B_{\pi^0}}$\,($\mathcal B_{K^0_S}$)
for the decay $D^0\to K^0_S\pi^0\eta^\prime$\,($D^+\to K^0_S\pi^+\eta^\prime$), and
${\mathcal B_{\eta^\prime}}$, ${\mathcal B_{\eta}}$, ${\mathcal B_{K^0_S}}$
and ${\mathcal B_{\pi^0}}$ denote the branching fractions of the decays
$\eta^\prime\to\pi^+\pi^-\eta$,
$\eta\to\gamma\gamma$, $K^0_S \to \pi^+\pi^-$, and
$\pi^0\to\gamma\gamma$, respectively, taken from the PDG~\cite{pdg2014}.
With the single-tag method, the CF decays $D^0(D^+)\to \bar K\pi\eta^\prime$ are indistinguishable
from the doubly Cabibbo-suppressed (DCS) decays $\bar D^0(D^+)\to \bar K(K)\pi\eta^\prime$.
However, the DCS contributions are expected to be small and negligible in the calculations of branching fractions,
but will be taken into account as a systematic uncertainty.

Taking $N_{D^{0}\bar{D}^{0}}=(10597\pm28_{\rm stat.}\pm 98_{\rm syst.})\times 10^3$
and
$N_{D^{+}D^{-}}=(8296\pm31_{\rm stat.}\pm65_{\rm syst.})\times 10^3$ from Ref.~\cite{bes3ddyield},
the branching fraction of each decay is determined with Eq.~(\ref{equ:branchingfraction})
and summarized in Table~\ref{tab:singletagN_MC}.

\begin{table*}[htp]
\centering
\caption{\label{tab:singletagN_MC}
$\Delta E$ requirements, input quantities and results for the determination of the branching fractions.
The efficiencies do not include the branching fractions for the decays of the daughter particles of
$\eta^\prime$, $\eta$, $K^0_S$, and $\pi^0$ mesons.
The uncertainties are statistical only.
}
\small
\begin{tabular}{lcccc} \hline
\multicolumn{1}{c}{Decay mode}& $\Delta E$ (MeV)  & $N_{\rm tag}$ &$\epsilon$ (\%) &$\mathcal B$ ($\times 10^{-3}$)
\\ \hline
$D^0\to K^-\pi^+\eta^\prime$   &$(-26,+28)$ &$2528\pm59$  &$10.97\pm0.08$&$6.43\pm0.15$\\
$D^0\to K^0_S\pi^0\eta^\prime$ &$(-35,+38)$ &$289 \pm26$  &$4.67\pm0.04$&$2.52\pm0.22$\\
$D^+\to K^0_S\pi^+\eta^\prime$ &$(-27,+28)$ &$267 \pm24$  &$7.23\pm0.05$&$1.90\pm0.17$\\ \hline
\end{tabular}
\end{table*}

\section{Systematic uncertainties}
\label{sec:sys}
The systematic uncertainties in the measurements of the branching fractions and the branching ratios, ${\mathcal R}^0\equiv\frac{{\mathcal B}(D^0\to K^0_S\pi^0\eta^\prime)}{{\mathcal B}(D^0\to K^-\pi^+\eta^\prime)}$, and ${\mathcal R}^+\equiv \frac{{\mathcal B}(D^+\to K^0_S\pi^+\eta^\prime)}{{\mathcal B}(D^0\to K^-\pi^+\eta^\prime)} $, are summarized in Table~\ref{tab:relsysuncertainties}. Each contribution, estimated relative to the measured branching fraction, is discussed below.

\begin{table*}[htp]
\centering
\caption{\label{tab:relsysuncertainties} Relative systematic uncertainties (in \%) in the branching fractions, ${\mathcal R}^0$, and ${\mathcal R}^+$.
The numbers before or after `/' in the last two columns denote the remaining systematic uncertainties of
${\mathcal B}(D^0\to K^{-}\pi^{+}\eta')$ and ${\mathcal B}(D^{0(+)}\to K^{0}_{S}\pi^{0(+)}\eta')$ that do not cancel
in the determination of ${\mathcal R}^{0}$ and ${\mathcal R}^{+}$.     }
\begin{small}
\begin{tabular}{lccccc}
\hline
Source & ${\mathcal B}(D^0\to K^{-}\pi^{+}\eta')$&${\mathcal B}(D^0\to K^{0}_{S}\pi^{0}\eta')$&${\mathcal B}(D^+\to K^{0}_{S}\pi^{+}\eta')$&${\mathcal
R}^0$&${\mathcal R}^+$\\
\hline
  Number of $D\bar D$ pairs
                                 & 1.0  & 1.0 & 0.9 & --/-- &1.0/0.9 \\
  Tracking of $K^\pm(\pi^\pm)$   & 3.0  & 2.0 & 2.5 &1.0/-- &1.0/--  \\
  PID of $K^\pm(\pi^\pm)$        & 2.0  & 1.0 & 1.5 &1.0/-- &0.5/--  \\
  $K_S^0$ reconstruction         &  --  & 1.5 & 1.5 & --/1.5& --/1.5 \\
  $\pi^0\,(\eta)$ reconstruction & 1.0  & 2.0 & 1.0 & --/1.0& --/--  \\
  $M_{\rm BC}$ fit               & 0.5  & 3.6 & 1.9 &0.5/3.6&0.5/1.9 \\
  $\eta^\prime$ mass window      & 1.0  & 1.0 & 1.0 & --/-- & --/--  \\
  $\Delta E$ requirement         & 0.1  & 2.4 & 4.5 &0.1/2.4&0.1/4.5 \\
  MC modeling                    & 1.6  & 0.5 & 1.7 &1.6/0.5&1.6/1.7 \\
  MC statistics                  & 0.7  & 0.9 & 0.7 &0.7/0.9&0.7/0.7 \\
  Quoted branching fractions     & 1.7  & 1.7 & 1.7 & --/0.1& --/0.1 \\
  $D^0\bar D^0$ mixing           & 0.1  & 0.1 & --  & --/-- & --/--  \\
  DCS contribution& 0.6  & 0.6 & 0.6 & --/-- & --/--  \\
  \hline
  Total & 4.8 & 6.0 & 6.6 & 5.3 & 6.0 \\
  \hline
\end{tabular}
\end{small}
\end{table*}

\begin{itemize}

\item
{\bf \boldmath Number of $D\bar D$ pairs}:
The total numbers of $D^0\bar D^0$ and $D^+D^-$ pairs produced in
the data sample are cited from a previous measurement~\cite{bes3ddyield}
that uses a combined analysis of
both single-tag and double-tag events in the same data sample.
The total uncertainty in the quoted number of $D^0\bar D^0~(D^+D^-)$ pairs
is 1.0\%~(0.9\%), obtained by adding both the statistical and systematic uncertainties in quadrature.

\item
{\bf \boldmath Tracking and PID of $K^\pm(\pi^\pm)$}:
The tracking and PID efficiencies for $K^\pm(\pi^\pm)$
are investigated using double-tag $D\bar D$ hadronic events.
A small difference between the efficiency
in the data sample and that in MC simulation~(called the data-MC difference) is found.
The momentum weighted data-MC differences
in the tracking [PID] efficiencies are determined to be
$(+2.4\pm0.4)\%$, $(+1.0\pm0.5)\%$, and $(+1.9\pm1.0)\%$
[$(-0.2\pm0.1)\%$, $(-0.1\pm0.1)\%$ and $(-0.2\pm0.1)\%$]
for $K^\pm$, $\pi^\pm_{\rm direct}$, and $\pi^\pm_{\rm in-direct}$,
respectively.
Here, the uncertainties are statistical
and the subscript $_{\rm direct}$ or $_{\rm in-direct}$ indicates
the $\pi^\pm$ produced in $D$ or $\eta^\prime$ decays, respectively.
In this work, the MC efficiencies have been corrected by the
momentum weighted data--MC differences in the $K^\pm(\pi^\pm)$ tracking and PID efficiencies.
Finally, a systematic uncertainty for charged particle tracking is assigned to be 1.0\%
per $\pi^\pm_{\rm in-direct}$ and 0.5\% per $K^\pm$ or $\pi^\pm_{\rm direct}$.
The systematic uncertainty for PID efficiency is taken as 0.5\% per $K^\pm$, $\pi^\pm_{\rm direct}$ or $\pi^\pm_{\rm in-direct}$.

\item
{\bf\boldmath $K_S^0$ reconstruction}:
The $K_{S}^{0}$ reconstruction efficiency,
which includes effects from the track reconstruction of the charged pion pair, vertex fit, decay length requirement and $K^0_S$ mass window,
has been studied with a control sample of
$J/\psi\to K^{*}(892)^{\mp}K^{\pm}$ and $J/\psi\to \phi K_S^{0}K^{\pm}\pi^{\mp}$~\cite{sysks}.
The associated systematic uncertainty is assigned as 1.5\% per $K^0_S$.

\item
{\bf \boldmath $\pi^0\,(\eta)$ reconstruction}:
The $\pi^0$ reconstruction efficiency,
which includes effects from the photon selection, 1-C kinematic fit and $\pi^0$ mass window,
is verified with double-tag $D\bar D$ hadronic decay samples
of $D^0\to K^-\pi^+$, $K^-\pi^+\pi^+\pi^-$ versus
$\bar D^0\to K^+\pi^-\pi^0$, $K^0_S\pi^0$~\cite{syspi0}.
A small data-MC difference in the $\pi^0$ reconstruction efficiency
is found.
The momentum weighted data-MC difference in $\pi^0$ reconstruction
efficiencies is found to be $(-0.5\pm1.0)\%$, where the uncertainty
is statistical.
After correcting the MC efficiencies by the momentum weighted data-MC
difference in $\pi^0$ reconstruction efficiency,
the systematic uncertainty due to $\pi^0$ reconstruction is assigned as 1.0\% per $\pi^0$.
The systematic uncertainty due to $\eta$ reconstruction is assumed to be the same as that for $\pi^0$ reconstruction.

\item
{\bf \boldmath $\eta^\prime$ mass window}:
The uncertainty due to the $\eta^\prime$ mass window is studied by fitting to the $\pi^+\pi^-\eta$ invariant mass spectrum of the $K^-\pi^+\eta^\prime$
candidates. The difference between the data and MC simulation in the efficiency of the $\eta^\prime$ mass window restriction is $(0.8\pm0.2)$\%.
The associated systematic uncertainty is assigned as 1.0\%.

\item
{\bf \boldmath $M_{\rm BC}$ fit}:
To estimate the systematic uncertainty due to the $M_{\rm BC}$ fit, we repeat
the measurements by varying the fit range [$(1.8415,1.8865)$\,GeV/$c^2$],
the signal shape\,(with different MC matching requirements)
and the endpoint\,(1.8865\,GeV/$c^2$) of the ARGUS function ($\pm0.2$\,MeV/$c^2$).
Summing the relative changes in the branching fractions for these three sources in quadrature
yields 0.5\%, 3.6\%, and 1.9\% for
$D^0\to K^-\pi^+\eta^\prime$, $D^0\to K^0_S\pi^0\eta^\prime$, and
$D^+\to K^0_S\pi^+\eta^\prime$, respectively,
which are assigned as systematic uncertainties.

\item
{\bf \boldmath $\Delta E$ requirement}:
To investigate the systematic uncertainty due to the $\Delta E$ requirement,
we repeat the measurements with alternative $\Delta E$
requirements of $3.0\sigma_{\Delta E}$ and $4.0\sigma_{\Delta E}$ around the fitted $\Delta E$ peaks.
The changes in the branching fractions, 0.1\%, 2.4\%, and 4.5\%, are taken as systematic uncertainties for
$D^0\to K^-\pi^+\eta^\prime$, $D^0\to K^0_S\pi^0\eta^\prime$, and
$D^+\to K^0_S\pi^+\eta^\prime$, respectively.

\item
{\bf MC modeling}:
The systematic uncertainty in the MC modeling is studied by varying
MC-simulated background sizes for the input
$M^{2}_{K\pi}$ and $M^{2}_{\pi\eta^\prime}$ distributions in the generator by~$\pm20\%$.
The largest changes in the detection efficiencies, 1.6\%, 0.5\%, and 1.7\% are taken as systematic uncertainties
for $D^0\to K^-\pi^+\eta^\prime$, $D^0\to K^0_S\pi^0\eta^\prime$, and
$D^+\to K^0_S\pi^+\eta^\prime$, respectively.

\item
{\bf MC statistics}:
The uncertainties due to the limited MC statistics are 0.7\%, 0.9\% and 0.7\%
for $D^0\to K^-\pi^+\eta^\prime$, $D^0\to K^0_S\pi^0\eta^\prime$, and
$D^+\to K^0_S\pi^+\eta^\prime$, respectively.

\item
{\bf Quoted branching fractions}:
The uncertainties of the quoted branching fractions for
$\eta^\prime \to \pi^+\pi^-\eta$,
$\eta\to \gamma\gamma$,
$K^0_S\to \pi^+\pi^-$, and
$\pi^0\to \gamma\gamma$ are taken from the world average and are 1.6\%,
0.5\%, 0.07\%, and 0.03\%~\cite{pdg2014},
respectively.

\item
{\bf $D^0\bar D^0$ mixing}:
Because $D^0\bar D^0$ meson pair is coherently produced in $\psi(3770)$ decay,
the effect of $D^0\bar D^0$ mixing on the branching fractions of neutral $D$
meson decays is expected to be due to the
next-to-leading-order of the $D^0\bar D^0$ mixing parameters $x$ and $y$~\cite{zzxing,asner}.
With $x=(0.32\pm0.14)\%$ and $y=(0.69^{+0.06}_{-0.07})\%$ from PDG~\cite{pdg2014}, we conservatively assign 0.1\% as the systematic uncertainty.

\item
{\bf DCS contribution}:
Based on the world-averaged values of the branching fractions, the branching fraction ratios between the known DCS decays and the corresponding CF decays are in the range of (0.2-0.6)\%. Therefore, we take the largest ratio 0.6\% as a conservative estimation of the systematic uncertainty of the DCS effects.

\end{itemize}

The above relative systematic uncertainties are added in quadrature, and a total
of 4.8\%, 6.0\%, 6.6\%, 5.3\% and 6.0\% for the measurements of
${\mathcal B}(D^0\to K^-\pi^+\eta^\prime)$,
${\mathcal B}(D^0\to K^0_S\pi^0\eta^\prime)$,
${\mathcal B}(D^+\to K^0_S\pi^+\eta^\prime)$,
${\mathcal R^0}$, and ${\mathcal R^+}$, respectively, is obtained.

\section{Summary and discussion}

Based on an analysis of an $e^+e^-$ data sample with an integrated luminosity of 2.93 fb$^{-1}$ collected at $\sqrt s= 3.773$~GeV
with the BESIII detector, we measure the branching fractions of hadronic $D$ meson decays to be:
${\mathcal B}(D^0\to K^-\pi^+\eta^\prime)=(6.43 \pm 0.15_{\rm stat.} \pm 0.31_{\rm syst.})\times 10^{-3}$,
${\mathcal B}(D^0\to K^0_S\pi^0\eta^\prime)=(2.52 \pm 0.22_{\rm stat.} \pm 0.15_{\rm syst.})\times 10^{-3}$, and
${\mathcal B}(D^+\to K^0_S\pi^+\eta^\prime)=(1.90 \pm0.17_{\rm stat.} \pm 0.13_{\rm syst.})\times 10^{-3}$.
The measured branching fraction of $D^0\to K^-\pi^+\eta^\prime$ is consistent with the previous result
measured by CLEO~\cite{prd48_4007,pdg2014}, but improved with a factor of 4 in precision.
The branching fractions of $D^0\to K^0_S\pi^0\eta^\prime$ and
$D^+\to K^0_S\pi^+\eta^\prime$ are determined for the first time.

Using the measured branching fractions, we determine the ratios of branching fractions to be
${\mathcal R^0}=0.39\pm0.03_{\rm stat.}\pm0.02_{\rm syst.}$ and
${\mathcal R^+}=0.30\pm0.03_{\rm stat.}\pm0.02_{\rm syst.}$.
${\mathcal R^0}$ agrees well with the value 0.4 predicted by the SIM,
but ${\mathcal R^+}$ significantly deviates from the expected value 0.9.
This deviation may arise from a possible phase difference between two isospin states in the SIM~\cite{lvcd}.
In our analysis, we do not find an obvious $K^{*}$ signal in the $K\pi$ invariant mass distributions, which is
consistent with the predictions of small $D^0\to \bar K^{*0}\eta^\prime$ and $D^+\to K^{*+}\eta^\prime$ contributions~\cite{Ddecay1,Ddecay2,Ddecay3}.

Summing over the branching fractions of $D\to \bar K\pi\eta^\prime$ decays and the other exclusive $D\to\eta^\prime X$ decays in PDG~\cite{pdg2014},
we obtain the sums of the branching fractions of all the exclusive $D^0\to\eta^\prime X$ and $D^+\to\eta^\prime X$ to be
$(3.23\pm0.13)\%$ and $(1.06\pm0.07)\%$, respectively.
They are consistent with the measured inclusive production ${\mathcal B} (D^0\to\eta^\prime X)=(2.48\pm0.27)\%$ and ${\mathcal B} (D^+\to\eta^\prime X)=(1.04\pm0.18)\%$~\cite{prd74_112005} within $2.5\sigma$ and $0.1\sigma$, respectively. This excludes the possibility of additional exclusive $D\to\eta^\prime X$ decay modes with large branching fractions.

\section{Acknowledgements}

The BESIII collaboration thanks the staff of BEPCII and the IHEP computing center for their strong support.
The authors are grateful to Fu-Sheng Yu, Jonathan L. Rosner, and Zhizhong Xing for helpful discussions.
This work is supported in part by National Key Basic Research Program of China under Contract No. 2015CB856700; National Natural Science Foundation of
China (NSFC) under Contracts Nos. 11335008, 11425524, 11475123, 11625523, 11635010, 11675200, 11735014, 11775230; the Chinese Academy of Sciences (CAS) Large-Scale Scientific Facility
Program; the CAS Center for Excellence in Particle Physics (CCEPP); Joint Large-Scale Scientific Facility Funds of the NSFC and CAS under Contracts Nos.
U1532257, U1532258, U1532101; CAS Key Research Program of Frontier Sciences under Contracts Nos. QYZDJ-SSW-SLH003, QYZDJ-SSW-SLH040; 100 Talents Program of CAS;
INPAC and Shanghai Key Laboratory for Particle Physics and Cosmology; German Research Foundation DFG under Contracts Nos. Collaborative Research Center
CRC 1044, FOR 2359; Istituto Nazionale di Fisica Nucleare, Italy; Koninklijke Nederlandse Akademie van Wetenschappen (KNAW) under Contract No. 530-4CDP03;
Ministry of Development of Turkey under Contract No. DPT2006K-120470; National Science and Technology fund; The Swedish Research Council; U. S. Department
of Energy under Contracts Nos. DE-FG02-05ER41374, DE-SC-0010118, DE-SC-0010504, DE-SC-0012069; University of Groningen (RuG) and the Helmholtzzentrum fuer
Schwerionenforschung GmbH (GSI), Darmstadt.

\end{document}